\documentclass[prl,twocolumn,showpacs,superscriptaddress,nofootinbib]{revtex4-2}
\pdfoutput=1
\usepackage{graphicx}
\usepackage{epstopdf}
\usepackage{bm}
\usepackage{amssymb}
\usepackage{color}
\usepackage{amsmath}
\usepackage{amstext}
\usepackage{latexsym}
\usepackage[usenames,dvipsnames]{xcolor}
\usepackage[colorlinks=true,citecolor=MidnightBlue,linkcolor=RubineRed,urlcolor=MidnightBlue]{hyperref}
\usepackage{float}

\def\x{{\rm\bf x}}

\newcommand{\beq}{\begin{equation}}
\newcommand{\eeq}{\end{equation}}
\newcommand{\beqa}{\begin{eqnarray}}
\newcommand{\eeqa}{\end{eqnarray}}

\usepackage{tikz,xcolor,hyperref}
\definecolor{lime}{HTML}{A6CE39}
\DeclareRobustCommand{\orcidicon}{
\begin{tikzpicture}
\draw[lime, fill=lime] (0,0)
circle[radius=0.16]
node[white]{{\fontfamily{qag}\selectfont \tiny \.{I}D}}; 
\end{tikzpicture}
\hspace{-2mm}
}
\foreach \x in {A, ..., Z}{%
\expandafter\xdef\csname orcid\x\endcsname{\noexpand\href{https://orcid.org/\csname orcidauthor\x\endcsname}{\noexpand\orcidicon}}
}

\begin{document}

\title{Quantum Turbulence Across Dimensions: Crossover from two- to three-dimension}

\author{Wei-Can Yang\hspace{-1.5mm}\orcidA{}
}\email{weicanyang@outlook.com}
\affiliation{Center for Gravitational Physics and Quantum Information, Yukawa Institute for Theoretical Physics, Kyoto University, Kyoto 606-8502, Japan}
\affiliation{Department of Physics, Osaka Metropolitan University, 3-3-138 Sugimoto, 558-8585 Osaka, Japan}

\author{Xin Wang\hspace{-1.5mm}}
\affiliation{School of Physics, Southeast University, Jiulonghu Campus, Nanjing 211189, China}

\author{Makoto Tsubota}\email{tsubota@omu.ac.jp}
\affiliation{Department of Physics, Osaka Metropolitan University, 3-3-138 Sugimoto, 558-8585 Osaka, Japan}
\affiliation{Nambu Yoichiro Institute of Theoretical and Experimental Physics (NITEP), Osaka Metropolitan University, 3-3-138 Sugimoto, Sumiyoshi-ku, Osaka 558-8585, Japan}

\begin{abstract}
We investigate the dynamic transition of quantum turbulence (QT) in a confined potential field, and by controlling the boundary aspect ratio, the system evolves from pure two-dimensional (2D) to quasi-two-dimensional, and finally to three-dimensional (3D). In the 2DQT, distinct Onsager vortex cluster formation and inverse energy cascade are observed, while 3DQT exhibits a direct energy cascade consistent with the Vinen turbulence decay rate, which display striking differences. By systematically altering the system height, we explore how dimensionality drives the differentiation of turbulence types and find that this transition is closely related to the excitation of Kelvin waves. Kelvin waves not only introduce additional dissipation mechanisms but also serve as mediators for direct energy transfer across scales. When the wavelength of the permitted Kelvin waves exceeds the critical size of vortex clusters, turbulence begins transitioning to 3D type, culminating in fully developed 3DQT at the characteristic scale. In the transitional region, we observe continuous variations in the decay rate and vortex cluster correlation functions.
\end{abstract}

\maketitle

Turbulence is a ubiquitous nonlinear and complex phenomenon in nature, characterized by energy transfer across multiple scales and fundamentally driven by multiscale interactions \cite{Davidson2015,Saffman1981,Frisch1995}. The origins of turbulence can be traced back to the 19th century, when its chaotic nature was first observed experimentally, while theoretical descriptions primarily rely on the Navier-Stokes equations to capture its dynamic behavior. A key distinction between 2D and 3D Turbulence lies in the directionality of energy cascades: 2D turbulence exhibits an anomalous inverse energy cascade, where energy is transferred from small-scale vortices to larger scales, ultimately forming large-scale vortex clusters. This phenomenon, first revealed by Kraichnan and others through theoretical and numerical studies, arises due to the conservation of vorticity and enstrophy in 2D flows \cite{Kraichnan_1967,Batchelor_1969}. In 2D turbulence, the enstrophy (the integral of the square of vorticity) cascades forward to smaller scales, while the conservation of total vorticity drives energy to accumulate at larger scales, leading to the formation of coherent vortex structures through the inverse cascade mechanism \cite{Paret_1999,Chen_2003}. In contrast, 3D turbulence exhibits a direct energy cascade, where energy is transferred from large scales to smaller scales and dissipated \cite{Richardson_1922,Jensen_1991}. Despite the significant differences between two- and three-dimensional turbulence,  the mechanisms governing the transition between them and their dynamic characteristics under gradually relaxed geometric constraints remain elusive, though some studies have explored energy cascade changes \cite{Smith_1996,Alexakis_2023}.

Quantum turbulence (QT), which is widely observed in Bose-Einstein condensates\cite{Abo_Shaeer_2001,Tsatsos_2016,Yang_2024_2} and superfluid $^4$He\cite{Vinen_2002,Skrbek_2021}, serves as the quantum counterpart of classical turbulence, offering a unique platform for studying turbulence dynamics. Unlike classical turbulence, QT consists of quantized vortices - topological defects characterized by fixed circulation and well-defined core structures \cite{Feynman_1955,Barenghi_2023,Tsubota_2013}. This quantization not only makes vortices highly identifiable and stable but also enables direct experimental visualization and precise theoretical modeling, eliminating the ambiguity associated with vorticity in classical systems and providing a clearer, more tractable framework for turbulence research.

In 3DQT, the dynamics are enriched by unique phenomena such as vortex reconnections and the excitation of Kelvin waves along vortex filaments. Vortex reconnections play a pivotal role in redistributing energy and generating cascades, while Kelvin waves mediate energy dissipation at smaller scales \cite{Minowa_2022,Tsubota_2000,Minowa_2025}. 
In contrast, 2DQT exhibits distinct behaviors due to its alignment with the ideal point-vortex model. A key feature of 2DQT is the spontaneous formation of Onsager vortex clusters—large-scale, coherent structures that arise from the inverse energy cascade \cite{Simula2014,Billam2014,Billam2015,Gauthier_2019,Han2019,Valani_2018,Reeves2013,Kanai2021,Gauthier_2019,Johnstone_2019,Groszek2016,Groszek2018,Yang_2024}. This clustering phenomenon is deeply rooted in statistical mechanics principles, where negative temperature states emerge as like-sign vortices aggregate into clusters while opposite-sign vortices annihilate \cite{Onsager_1949}. The inverse energy cascade observed in 2DQT highlights the universality of this feature across classical and quantum systems, yet only in QT can the actual dynamics of individual vortices be directly observed and studied. The consistency between quantum and classical turbulence in the dimension energy cascade provides a valuable framework for exploring the dimension crossing in the study of turbulence \cite{Muller_2020}.

In this Letter, we investigate the transition of QT from 2D to 3D, with the dimensional crossover driven by varying the system’s aspect ratio. By imprinting the same number of quantized vortices in traps with different aspect ratios and allowing them to evolve freely, we observe how vortex dynamics exhibit characteristic features of different dimensional regimes. Through a comprehensive analysis of vortex clustering, decay dynamics, and energy spectra, we identify the key role played by Kelvin waves in mediating the transition between 2D and 3D QT.

To study QT in a controlled framework, we adopt the dimensionless form of the Gross-Pitaevskii (GP) equation. The original GP equation is
\begin{equation}
    i\hbar \frac{\partial \psi(\mathbf{r}, t)}{\partial t} = \left[ -\frac{\hbar^2}{2M} \nabla^2 + g |\psi(\mathbf{r}, t)|^2 - \mu +V_{trap} \right] \psi(\mathbf{r}, t)
\end{equation}
Here, \(\psi(\mathbf{r}, t)\) represents the condensate wavefunction, while \(\hbar\), \(M\), \(g\), \(\mu\), and \(V_{\mathrm{trap}}\) are the reduced Planck constant, particle mass, coupling constant, chemical potential, and external trapping potential, respectively.

\begin{figure}[t]
    \centering
   \includegraphics[width=13cm,trim=60 50 0 0]{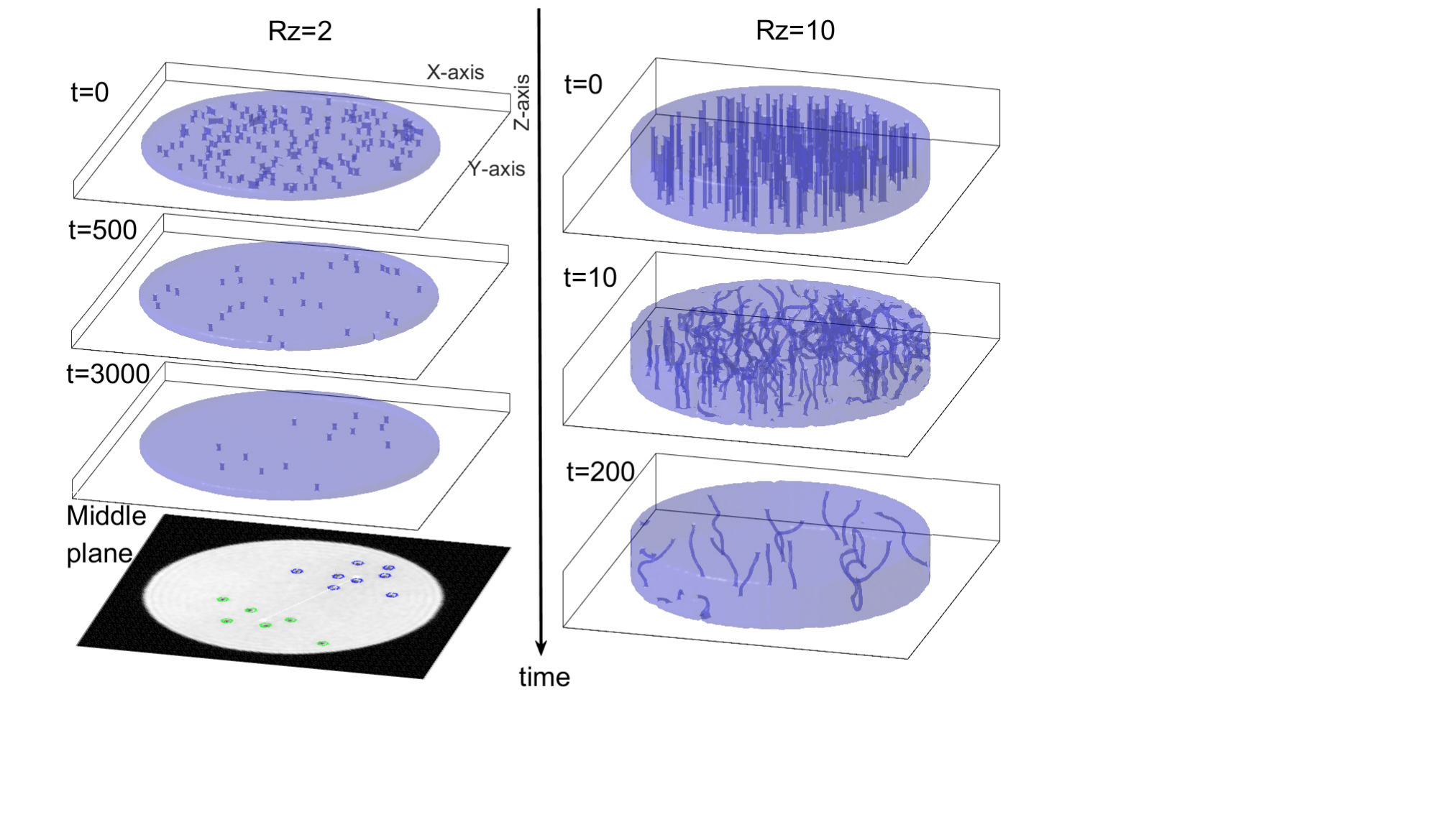}  \caption{Density evolution of vortex dynamics for different potential trap heights \(R_z\). The left column shows the evolution in the quasi-2D regime with \(R_z = 2\), the final figure presents a cross-sectional view after the formation of stable vortex clusters, where positive and negative vortices are represented in green and purple, respectively. The right column shows the evolution in the 3D regime with \(R_z = 10\). Movie can find in supplementary \cite{Movie}.}
 \label{figure1}
\end{figure}

We introduce dimensionless variables by defining \(\mathbf{r} = \mathbf{r}' l_0\) with characteristic length \(l_0\), \(\psi(\mathbf{r}, t) = \sqrt{\rho_0} u(\mathbf{r}', t')\) with density scale \(\rho_0 = \mu/g\), and \(t = t' t_0\) with time scale \(t_0 = Ml_0^2/\hbar\). Substituting these into the GP equation, we obtain the dimensionless form:
\begin{equation}
    i\frac{\partial u(\mathbf{r}', t')}{\partial t'} = \left[ -\frac{1}{2} \nabla'^2  + \mu' (|u(\mathbf{r}', t')|^2 - 1) + V'_{trap} \right] u(\mathbf{r}', t')
\end{equation}
where \(\mu' = \frac{\mu l_0^2 M}{\hbar^2}\). The dimensionless coherence length is \(\xi' = \xi/l_0 = \sqrt{1/(2\mu')}\). In this study, we choose \(\mu' = 1.4\), then \(\xi' = 0.6\) to balance numerical efficiency and physical relevance.

To confine the dynamics within a controlled region, we adopt a cylindrical box potential field defined as
\begin{equation}
    V'_{trap}(\mathbf{r'}) = 
    \begin{cases} 
      0, & \text{if } \sqrt{x^2 + y^2} \leq R \text{ and } 0 \leq z \leq R_z, \\ 
      100, & \text{otherwise}.
    \end{cases}
\end{equation}
This potential confines the system within a cylinder of radius \(R\) and height \(R_z\), effectively restricting the motion of quantized vortices and ensuring that the boundary conditions align with realistic experimental settings. The high potential barrier outside the cylindrical region prevents particles from escaping, thereby accurately mimicking the hard-wall boundaries commonly used in both experimental setups and previous studies on vortex dynamics \cite{Simula2014,Kanai2021}.

\begin{figure}[t]
    \centering
   \includegraphics[width=8.2cm,trim=30 30 0 0]{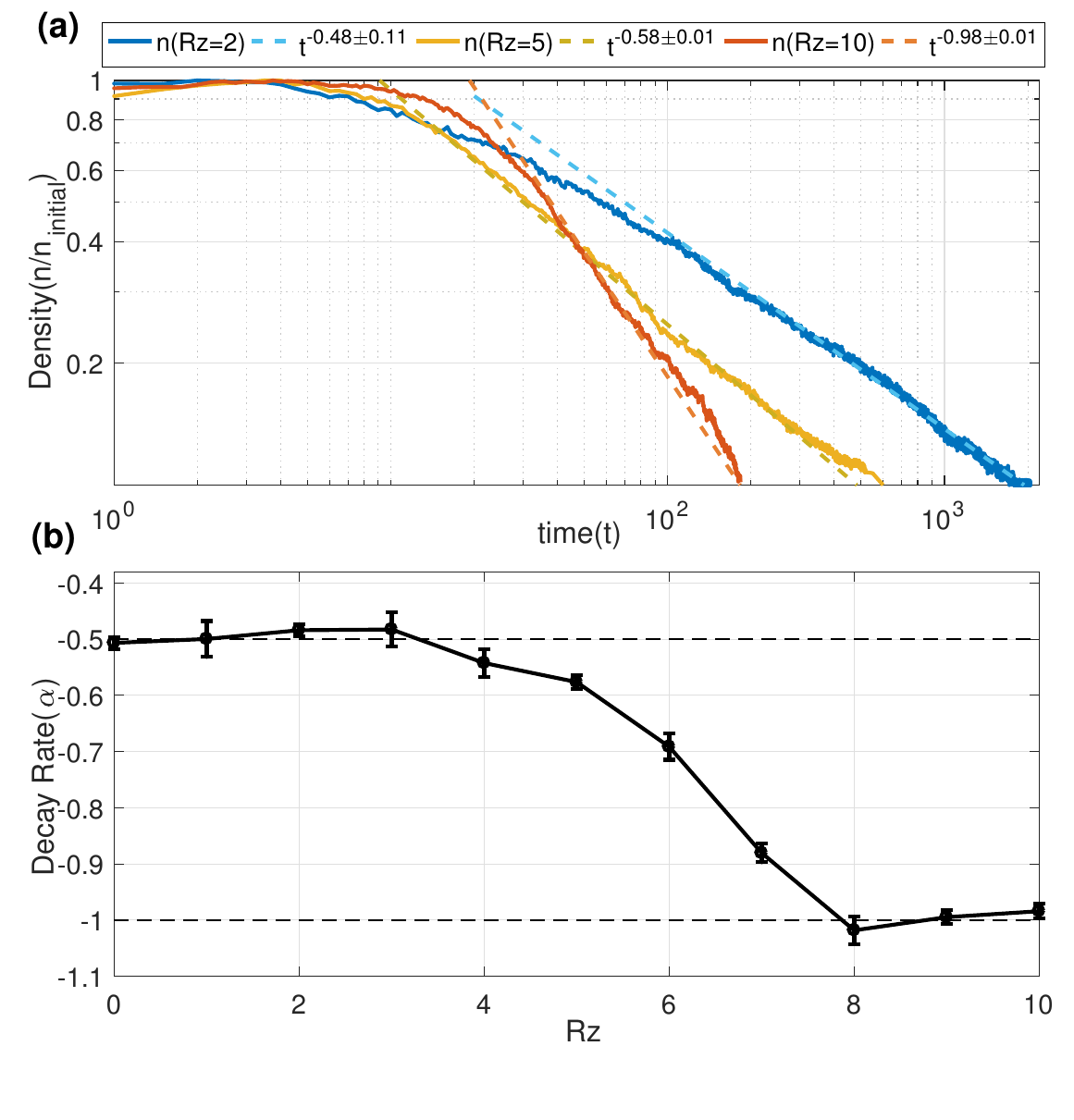}
    \caption{(a) Vortex decay process (solid) and fitting line (dash) for $R_z=2$ (blue), $R_z=5$ (yellow) and $R_z=10$ (red) case. (b) Decay rate versus height.}
    \label{decayrate}
\end{figure}

As shown in Fig.~\ref{figure1}, by using this boundary condition, we fixed the radius $R=30$ of the cylinder, thus simulating the transition from quasi-2D to 3D vortex dynamics by changing the height $R_z$. Initially, 100 positive and 100 negative vortices are randomly placed using the vortex imprinting method, and the system is allowed to undergo relaxation dynamics \cite{Yang_2024}.

In the left half of Fig.~\ref{figure1}, where a small height $R_z = 2$ is selected, the turbulence dynamic clearly exhibits two-dimensional characteristics. Vortices can be approximated as point vortices, and vortex decay occurs only through vortex-antivortex annihilation. After extensive annihilation, two large vortex clusters, known as Onsager vortices, emerge and keep stable, consistent with experimental observations \cite{Gauthier_2019}. The corresponding Movie can be found in the supplementary\cite{Movie}.

In contrast, the right half of Fig.~\ref{figure1} shows the system’s evolution at a larger height $R_z = 10$. In this case, vortex lines quickly twist and entangle, forming the characteristic features of 3DQT. Driven by vortex reconnections and Kelvin waves, vortex decay occurs rapidly, preventing the formation of coherent vortex clusters, highlighting the suppression of Onsager vortex formation in the 3D regime.

\begin{figure}[t]
    \centering
   \includegraphics[width=8.5cm,trim=30 0 0 0]{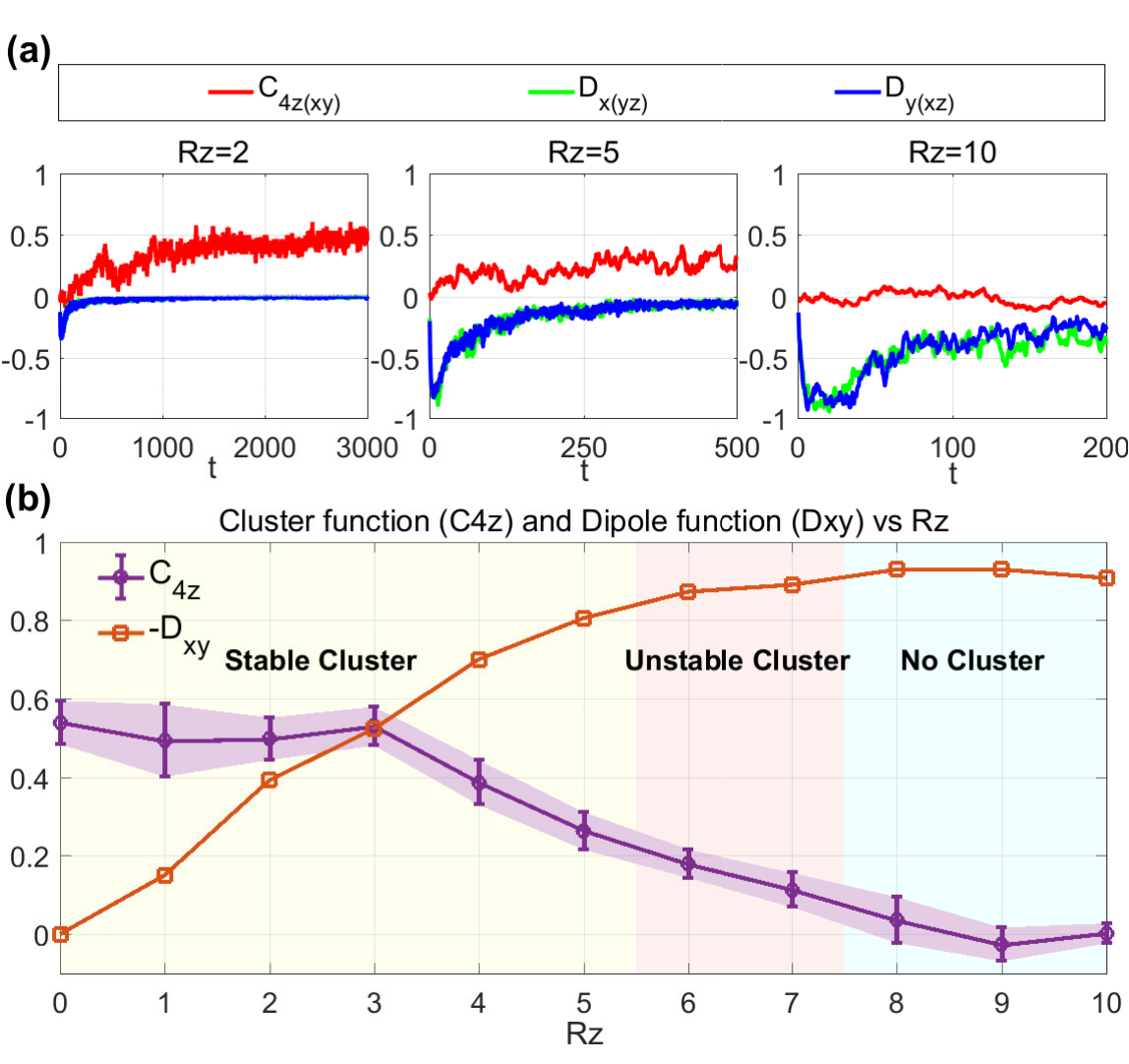}
    \caption{(a) Vortex cluster (red line) and dipole (blue and green line) correlation function evolution in $R_z=2,5,10$ case. (b) Cluster (purple) and dipole (orange) correlation function versus height and phase diagram. Yellow phase means stable cluster formation, pink phase means cluster can form but not stable forever, blue phase means no cluster formation.}
    \label{cluster}
\end{figure}

Clearly, as the height of the potential trap increases, the system undergoes a transition from quasi-2D to 3D QT. More precisely, this corresponds to a transition from an inverse energy cascade to a direct energy cascade. To investigate the transition behavior and the underlying mechanisms, we conduct simulations at multiple intermediate heights, tracking the distinct transition behaviors and the emergence of key dynamical processes during the evolution. And each set of data is the average of more than 20 simulations.

From the purely 2D case at \(R_z = 0\) to the fully 3D case at \(R_z = 10\), we select $11$ different heights at intervals of 1 to simulate the system’s evolution. Starting from the initial vortex configurations and allowing the system to relax to a fully stable state, we systematically track the QT dynamics as the system size varies. 
Fig.~\ref{decayrate} shows the vortex decay process characterized by \(n \sim t^\alpha\), where \(n\) is the vortex density and \(\alpha\) is the decay exponent. For \(R_z \leq 3\), the system exhibits a decay behavior consistent with the 2D three-vortex-process decay mechanism, with \(\alpha = -1/2\) \cite{Baggaley_2018,Andrew_2020}. In contrast, for \(R_z \geq 8\), the decay follows the characteristic behavior of 3D Vinen turbulence, with a faster decay rate $\alpha=-1$ \cite{Baggaley_2012,Vinen_2001,Cidrim_2017}. 
Between \(R_z = 3\) and \(R_z = 8\), the decay exponent \(\alpha\) shows a continuous transition, indicating the coexistence of both 2D and 3D QT effects within this intermediate region. This crossover reflects the gradual shift in the dominant dynamical processes, where 2D-like vortex-antivortex annihilation coexists with 3D vortex reconnections and the excitation of Kelvin waves.

To deeper explore the underlying dynamical mechanisms, we investigate two key functions: the vortex cluster correlation function and the dipole correlation function, which represent the formation proportions of vortex clusters and vortex dipoles, respectively. 

\begin{figure}[t]
    \centering
   \includegraphics[width=5.5cm,trim=60 50 0 10]{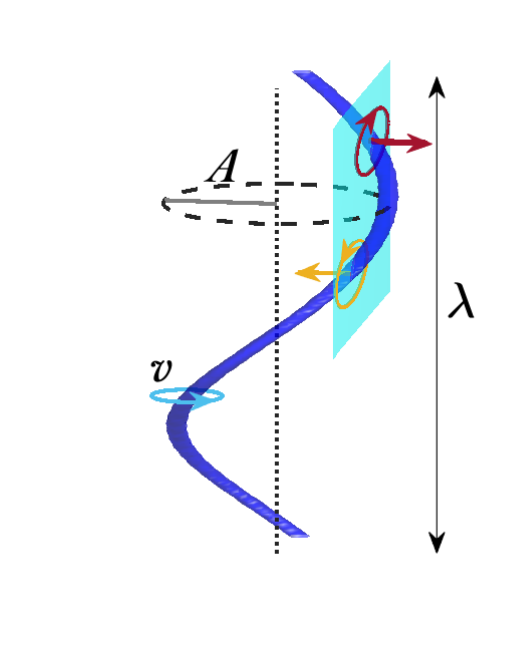}
    \caption{Schematic representation of a Kelvin wave along a vortex line. The blue square represents a side-cutting plane, where a pair of positive and negative vortex cores can be observed due to the local twisting induced by the Kelvin wave.}
    \label{kelvinwave}
\end{figure}

The cluster correlation function, denoted as \(C_B\), measures the fraction of vortices that form coherent clusters based on the proportion of same-sign vortices among the nearest \(B\) neighbors. Specifically, for the case of \(B = 4\), the 4th-nearest-neighbor cluster correlation function \(C_4\) is given by \cite{White_2012,Reeves2013}
\begin{equation}
    C_4 = \frac{1}{N} \sum_{i=1}^{N} \sum_{j=1}^{4} \frac{c_{ij}}{4},
\end{equation}
where \(N\) is the total number of vortices, and \(c_{ij} = 1\) if the \(j\)-th nearest vortex to vortex \(i\) has the same circulation sign, and \(c_{ij} = -1\) otherwise. The value of \(C_B\) reflects the degree of cluster formation, with higher values indicating a stronger tendency for same-sign vortices to aggregate.

The dipole correlation function, denoted as \(D\), captures the proportion of neighboring vortex pairs with opposite circulations, representing the degree of vortex dipole formation. It is calculated as
\begin{equation}
    D = \frac{1}{N} \sum_{i=1}^{N} d_i,
\end{equation}
where \(d_i = -1\) if the closest vortex to vortex \(i\) has the opposite circulation sign, and \(d_i = 1\) otherwise. A larger negative value of \(D\) indicates a strong tendency for vortex-antivortex pairs to form and dominate the dynamics.

Due to the anisotropic nature of quasi-2D QT along the \(z\)-direction and the isotropy of 3DQT, we consider the cluster correlation function along the \(z\)-direction and the dipole correlation function in the \(x\) and \(y\) directions. As shown in Fig.~\ref{cluster}(a), during the evolution at \(R_z = 2\) in the quasi-2D regime, the cluster correlation function gradually increases and eventually stabilizes at a large positive value $C_{4z}\approx0.5$, indicating the formation and stability of Onsager vortex. When the height increases to \(R_z = 5\), although the result is less pronounced, stable vortex clusters still persist. However, at \(R_z = 10\), the cluster correlation function is always zero, indicating the complete absence of vortex clusters.

The clustering correlation function effectively characterizes the dynamic evolution of QT, while the dipole correlation function provides further insights. As the height increases from small to large, the initial values of the dipole correlation function in the \( x \) and \( y \) directions emerge and grow, which corresponds to the presence of Kelvin waves. This can be explained using Fig. \ref{kelvinwave}, where the existence of Kelvin waves leads to the appearance of dipole structures observable in the side cross-section of the vortex line.

In Fig.~\ref{cluster}(b), we plot the height dependence of the cluster and dipole correlation functions, along with the corresponding phase diagram. As the height increases, the cluster correlation function begins to decrease at \(R_z \geq 3\), which is consistent with the dimensional change of the decay rate, indicating that the inverse energy cascade of the two-dimensional features begins to be affected at the same time as the three-dimensional decay process occurs. At this stage, Kelvin wave dynamics gradually become dominant. When the height reaches \(R_z \geq 6\), tthe dipole correlation function approaches its maximum, signifying that the initial Kelvin wave density has peaked, after which vortex clusters lose their stability.  Finally, for \(R_z > 7\), no vortex clusters are observed, the system fully transitions to the 3DQT regime.
This process highlights the critical role of Kelvin waves in driving the dimensional transition of QT, marking the shift from cluster-dominated dynamics in the 2D regime to reconnection- and wave-dominated dynamics in the 3D regime.

\begin{figure}[t]
    \centering
   \includegraphics[width=9cm,trim=0 20 0 10]{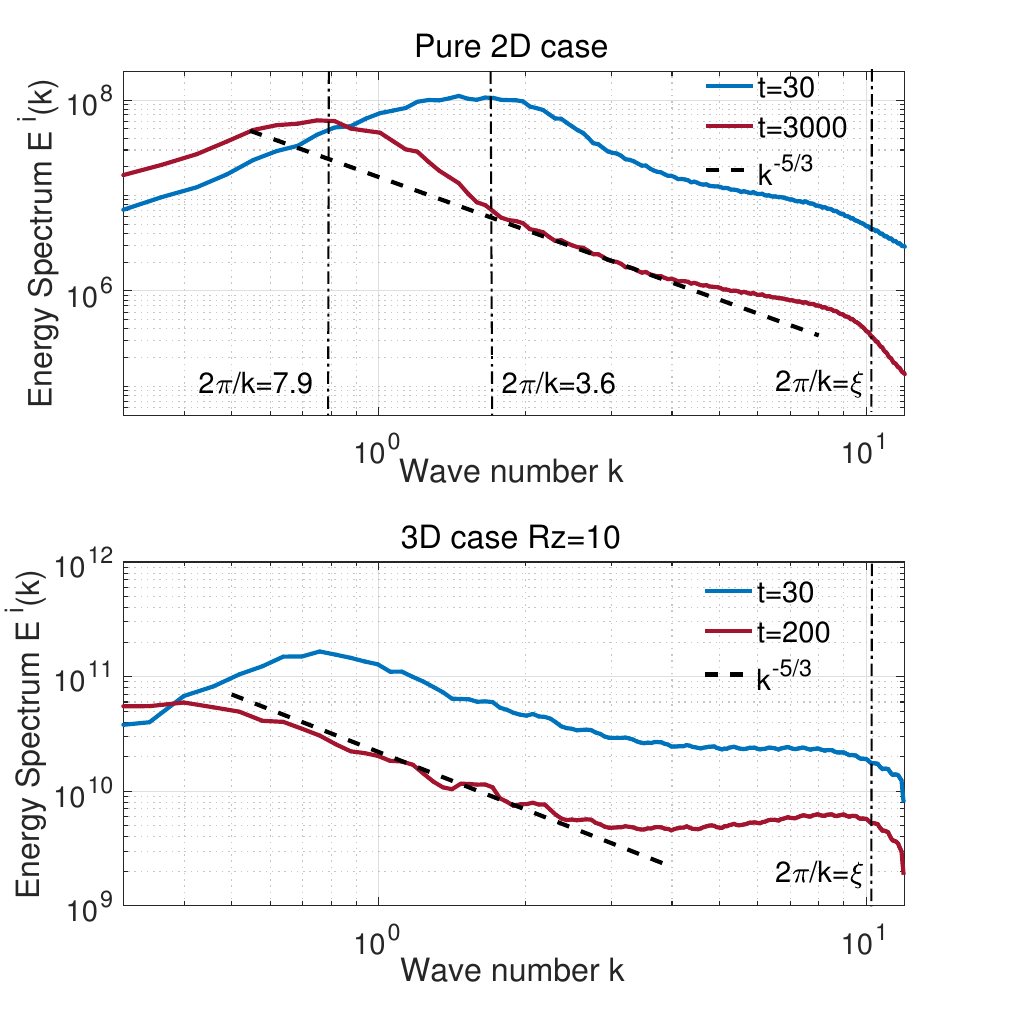}
\caption{Energy spectra for the pure 2D case (upper panel) and the 3D case (lower panel). The blue line represents the initial state, while the red line corresponds to the final stable state. In the pure 2D case, the three dotted vertical lines represent the characteristic scales of the system: the maximum energy peak scale, the minimum cluster scale, and the vortex scale. The diagonal dashed lines correspond to the Kolmogorov spectrum.}
\label{spectrum}
\end{figure}

To further confirm the crucial role of Kelvin waves in the energy cascade, we calculate the quantum kinetic energy spectrum for the incompressible part \cite{Reeves_2014,Michikazu_2005}:
\begin{eqnarray}\label{energyspectrum}
E_k^{i}(k) = \frac{1}{2} \int_{k \leq |\mathbf{k}'| < k+\Delta k} \left| \mathbf{V}^i(\mathbf{k}')\right|^2 d^3k'
\end{eqnarray}
where the incompressible field in Fourier space is given by
$\mathbf{V}^i(\mathbf{k}) = \mathcal{F}[\mathbf{V}] - (\mathbf{k} \cdot \mathcal{F}[\mathbf{V}])\mathbf{k}/|\mathbf{k}|^2$, here
$\mathbf{V} = \psi \mathbf{u}$ and 
the superfluid velocity $\mathbf{u}$ defined as
$\mathbf{u} = \frac{i}{2} (\psi^* \nabla \psi - \psi \nabla \psi^*)/|\psi|^2$.

As shown in Fig. \ref{spectrum}, in the pure 2D QT state, the energy spectrum exhibits a clear inverse energy cascade at low wavenumbers. Additionally, an obvious energy peak emerges above the Kolmogorov law in the inertial range, which corresponds to the formation of vortex clusters, indicating the accumulation of energy at low wave numbers \cite{Simula2014}. Surprisingly, the scale of the energy peak is closely related to the critical height associated with the dimensional transition. Specifically, when the maximum possible Kelvin wavelength allowed by the system height satisfies
\begin{equation}
 R_z = \lambda_{\mathrm{Kelvin}} \geq (2\pi/k)_{\mathrm{min}} \approx 3.6   
\end{equation}
 the Kelvin waves start to introduce a directly energy cascade that interferes with the inverse cascade characteristic of 2DQT. This interference is reflected in the decrease of the decay rate and the cluster correlation function, as observed in Fig.~\ref{decayrate} and Fig.~\ref{cluster}. Eventually, when the system height exceeds \(R_z > (2\pi/k)_{\mathrm{peak\ max}} \approx 7.9\), the inverse energy cascade completely vanishes, and no vortex clusters form, indicating a full transition to the 3DQT regime. This transition signifies that the dynamics of the Kelvin wave cascade dominate the system, effectively suppressing large-scale vortex clustering and driving the cascade processes characteristic of 3DQT.

In summary, by comparing the vortex relaxation dynamics at different potential trap heights, we have quantitatively studied, for the first time, the transition of QT behavior and energy cascades from two dimensions to three dimensions. The vortex decay rate, cluster correlation function, and energy spectrum demonstrate remarkable consistency throughout the transition. 
We conclude that the competition between the Kelvin wavelength and the characteristic vortex cluster size determines the dominance of 2D or 3D features. Only when the Kelvin wavelength exceeds the characteristic vortex cluster size do vortex reconnections and Kelvin-wave-mediated forward energy cascades begin to dominate. This leads to the disruption of large-scale structures typical of 2DQT and facilitates energy dissipation in the 3D regime, marking the onset of the dimensional transition.
Kelvin waves, corresponding to the twisting of vortices in the transverse direction, play a crucial role in breaking vorticity conservation. This provides direct evidence that the conservation of vorticity is the key factor determining the direction of energy cascades. The breaking of this conservation allows for the directly cascade to emerge, ultimately driving the system into a fully developed 3DQT state.
Experimentally, well-established techniques are already available to validate our results by systematically comparing the vortex dynamics at different potential trap heights\cite{Gauthier_2019}.

{\it Acknowledgements.}
We would like to thank Nir Navon for valuable discussions.
M. T. acknowledges the support from JSPS KAKENHI Grant Number JP23K03305 and JP22H05139.

\end{document}